\documentclass[referee]{raa}            % referee version: for submission

\usepackage{graphicx,times}             %for PS/EPS graphics inclusion, new
\usepackage{lscape}

\begin{document}

% Misc definitions
\newcommand{\kms}{km~s$^{-1}$}
\newcommand\sk[2] {Sk\,{$-#1{^\circ}#2$}}
\newcommand{\tnc}{\,\tablenotemark{c}}
\newcommand{\tnd}{\,\tablenotemark{d}}

\newcommand{\CIV}{{C}{IV}}    %  CIV
\newcommand{\MgII}{{Mg}{II}}    %  MgII
\newcommand{\OII}{{O}{II}}    %  CIV
\newcommand{\OIII}{{O}{III}}    %  MgII
\newcommand{\Hb}{H$\beta$}    %  Hbeta
\newcommand{\Ha}{H$\alpha$}   % Halpha

\newcommand{\ks}{\mbox{$K_\mathrm{s}$}}

\newcommand{\degpp}{\mbox{$^\circ\mskip-7.0mu\,$}}
\newcommand{\degpoint}{\mbox{$^\circ\mskip-7.0mu.\,$}}
\newcommand{\secpp}{\mbox{$''\mskip-7.6mu\,$}}
\newcommand{\secpoint}{\mbox{$''\mskip-7.6mu.\,$}}
\newcommand{\lya}{Lyman~$\alpha\,\,$}
\newcommand{\rAA}{{\AA \enskip}}
\def\arcs{$''~$}

   \title{High resolution deep imaging of a bright radio quiet QSO at $z\sim 3$
%\footnotetext{$*$ Supported by the National Natural Science Foundation of China.}
}
%   \subtitle{I. Place Your Subtitle Here}

   \volnopage{Vol.0 (200x) No.0, 000--000}      %%preserved for Editor. DOn't remove!
   \setcounter{page}{1}          %%starting page, preserved for Editor. DOn't remove!

   \author{Yiping Wang
      \inst{1}
   \and Wei He
      \inst{1}
   \and Toru Yamada
      \inst{2}
   \and Ichi Tanaka
      \inst{3}
   \and Masanori Iye
      \inst{4}
   \and Tuo Ji
      \inst{5}
   }

   \institute{Key Laboratory of Optical Astronomy, National Astronomical
observatories, Chinese Academy of Sciences, Beijing 100012, China; {\it ypwang@bao.ac.cn}\\
        \and
             Tohoku University, Aramaki, Aoba, Sendai 980-8578, Japan\\
        \and
             Subaru telescope, National Astronomical Obs. of Japan\\
        \and
             National Astronomical Obs. of Japan, Mitaka, Japan\\
        \and
             Antarctic Astronomy Division, Polar Research Institute of China\\
   }

   \date{Received~~2009 month day; accepted~~2009~~month day}

\abstract{We have obtained deep ${\it J \& Ks}$-band images centered on a bright radio quiet QSO UM402 ($z_{em}=2.856$) using IRCS camera and AO systems on Subaru Telescope, as well as retrieved WFC3/F140W archive images. A faint galaxy ($m_{k}=23.32\pm 0.05\,\,$ in the Vega system) that lies $\sim 2\secpoint 4$ north of the QSO sightline has been clearly resolved in all three deep high resolution datasets, and appears as an irregular galaxy with two close components in the ${\it Ks}$-band images (separation $\sim 0\secpoint 3$). Given the small impact parameter ($b=19.6\,\,{\rm kpc}$, at $z_{lls}=2.531$), as well as the red color of $(J-Ks)_{vega}\sim 1.6$, it might be a candidate galaxy giving rise to the Lyman Limit system absorption at $z_{abs}=2.531$ seen in the QSO spectrum. After carefully subtracting the PSF from the QSO images, the host galaxy of this bright radio quiet QSO at $z\sim 3$ was marginally revealled. We placed a low limit of the host component of $m_{k}\sim 23.3$ according to our analyses.
}

   \authorrunning{Y. P. Wang et al.}            %author_head in even pages
   \titlerunning{AO deep imaging of QSO UM402 field }  % title_head in odd pages

   \maketitle
%% The author head (on even pages) and the title head (on odd pages) will be
%% automatically extracted from \author{} and \title{}. Whenever the title is too long,
%% you will be asked to supply a shorter one by inserting either \authorrunning{} or
%% \titlerunning{} before \maketitle. Anyway, you can specify your own heads.
%%
%%
%% Note: In the following text body of your manuscript, please note several differences from
%%       other major journals:
%% (1) \subsection{Please Capitalize the First Letter of Each Notional Word in Subsection Title}
%% (2) Please Capitalize the First Letter of Each Notional Word in all tables' captions

%
%________________________________________________ sections below
%
\section{Introduction}           %% first-level sections will be auto-capitalized

Recent cosmological simulations have made significant progress on the understanding of physical properties of quasar absorption line systems and their implications for the ionizing background. Lyman Limit systems (LLSs) as a member of quasar absorption line (QAL) family, are defined to be optically thick at the Lyman limit($\lambda < 912{\rm \AA}$) and of a neutral hydrogen column density $N(H{\sc I})> 10^{17} {\rm cm}^{-2}$, attract much attention within these years. The numerical simulations by Fumagalli et al. (2011) suggest that cold gas streams of circumgalactic medium mostly appear as LLSs ($N_{HI}> 10^{17} cm^{-2}$), while gas clumps in the streams give rise to DLAs ($N_{H{\sc I}}> 10^{20.3} cm^{-2}$) as well. Meanwhile, the simulations show that cold streams are unlikely to produce the large equivalent widths of low-ion metal absorption lines, indicating they may arise from outflows.

Although sophisticated numerical simulations with much higher resolution are available these years, the questions on the physical origin of the absorption lines, i.e. whether the absorbing gas primarily traces the cool extended regions of dark matter halos, dwarf satellite galaxies, galactic discs or supernovae-driven outflows, remain still in dispute (Katz et al. 1996, Kohler \& Gnedin 2007, Prochaska et al. 2010, Dekel et al. 2009, Fumagalli et al. 2011, Rahmati \& Schaye 2014). For example, Erkal (2014) studied LLSs using simulations which had higher spatial and mass resolution than that of Kohler \& Gnedin (2007), and concluded that the majority of LLSs reside in low-mass halos, opposite to what Kohler \& Gnedin (2007) have found. However, we understand that the treatment of the complicated physical processes in the simulation would cause uncertainties, and observational evidences and good statistics would set constraints to these models in this case.

Deep imaging and spectroscopy of close QSO/galaxy pairs would provide unique opportunity to study both the gas flows around galaxies and their stellar population, as well as to determine the morphology and orientation of the galaxy disks in space. By performing a kinematic comparison of these absorber galaxies and the QSO aborption line systems, we hopefully could effectively investigate the matter of circumgalactic medium (CGM) which are accreting onto galaxies or being expelled by galactic wind. Therefore, identifying LLS absorbers using high resolution deep imaging would be a crucial first step of the study, and there has already been a definite improvement nowadays on the detection power of faint intervening galaxies in the QSO sightline both from space and AO-assisted ground-based large facilities, especially towards the peak epoch of galaxy formation ($z\sim 3$). Recent work on nearby Lyman Limit Systems shows the example and the importance to extend such a study to the under-exploited high-z universe, with a feast of galaxy formation activities (Stocke et al. 2010, Ribaudo et al. 2011).

On the other hand, the studies on the high-z quasar hosts have also received increasing attention lately, since they open an important avenue to study the assembly and evolution of massive galaxies, in particular in relation to the growth of their central black holes. Although deblending the host components of high-z QSOs from the bright central AGNs is non-trivial, previous efforts on this subject have made significant progress depending on high resolution images from space(HST) and AO techniques of ground-based large facilities. A sample of faint or medium-bright QSOs, including RLQs and RQQs at $z\sim 2-3$, have been observed especially with the NICMOS camera on HST (Ridgway et al.2001, Kukula et al.2001, Peng et al. 2006). Moreover, ground-based 8m class telescopes, especially with adaptive optics, offer a high spatial resolution for a powerful detection of high-z QSO host galaxies (Falomo et al. 2008, Schramm et al. 2008, Wang et al.2013). However, host properties from the studies on current sample have been the subject of some debate. Except that Falomo et al.(2008) resolved a large host galaxy of a medium-bright RLQ at $z\sim 2.9$, we lack evidences towards the peak epoch of galaxy formation at $z\sim 3$, especially for luminous QSOs.

{\bf UM402} is a bright, high redshift radio-quiet QSO discovered by Macalpine \& Lewis (1978). It is bright enough to permit detailed spectroscopic observations with a resolution ranging from a few to hundreds $\rm km \, s^{-1}$, and showing clearly the strong and
sharp \lya  and $\rm CIV$ emission lines, as well as a Lyman Limit
system at $z=2.531$ ($N_{HI}>4.6\times10^{17}cm^{2}$) even with a low-resolution spectrum (Sargent et al. 1989). Previous deep imaging in the optical band of this QSO field reported the detection of several close neighbors ($\theta \sim 4\secpoint 7 - 7''$), which are all spectroscopically confirmed at redshift $z<1$ (Le Brun et al. 1993, Guillenmin \& Bergeron 1997). Thus, the galaxy counterpart of the LLS seen in the QSO spectrum is still not identified, indicating it might be much fainter than the previous detection limit of $m_{r}(3\sigma_{sky})=25.2$, or much closer to the QSO sightline.

{\bf UM402} again is one of the high-z QSOs which we selected for a pilot study, using IRCS camera and AO system of Subaru telescope. Several issues were carefully considered during the target selection, 1) high-z QSOs near the era of peak QSO activity and cosmic star formation history at $z\sim 2-3$, are specially selected due to their importance on the understanding of the galaxy formation scenario; 2) there should be a bright guide star ($R<15$) sufficiently close to the QSO sightline ($<30''$), in order to be observed with IRCS+AO36 on Subaru telescope; 3) the emission lines \Ha(6563\rAA), \Hb(4861\rAA), \OII(3727\rAA) and \OIII(5007\rAA) should be avoided to be included in the observing bands. This is important for the host mass estimation and the host continuum property studies; 4) there exists a suitable PSF calibration star for the PSF subtraction from the QSO images. We will elaborate on this point in section 2.1.

In this paper, we present the initial results from the deep {\it $J\& Ks$}-band images centered on QSO UM402 ($z_{em}=2.856$) using IRCS camera and AO systems on Subaru Telescope, as well as the WFC3/F140W archived images (PI: Dawn Erb, HST proposal ID 12471). The cosmological parameters $\Omega=0.27,\,\,\Lambda=0.73$ and $\rm H=71\,km/s/Mpc$ are adopted throughout. 

%% Authors can give a citation as 'Michel et al. 1992'.
%% You may also use \cite, \citep and \citet for citation, and use Table~1 or Figure~1
%% and so forth. Using \ref and \label for cross-references of Tables/Figures
%% is a good way in adjusting/adding/removing text, tables or figures.

\section{Observations}
\subsection{Selection of AO guide star and PSF calibration star}
For Subaru AO36 adaptive optics systems, a natural guide star(NGS) sufficiently close to the sightline of the target is required as a reference source to assess the degradation of the wavefronts due to the turbulent atmosphere. There is a bright star with magnitude of $R=13.8$ and an angle distance to UM402 of $\theta_{GS}\sim 31''$. We selected this bright star as our AO guide star, and expected to obtain the AO corrected PSF better than $0\secpoint 2$ if the natural seeing is $<0\secpoint 6$ (Takami et al. 2004).

\begin{figure}[h]
  \begin{center}
   \includegraphics[width=59mm,height=58mm]{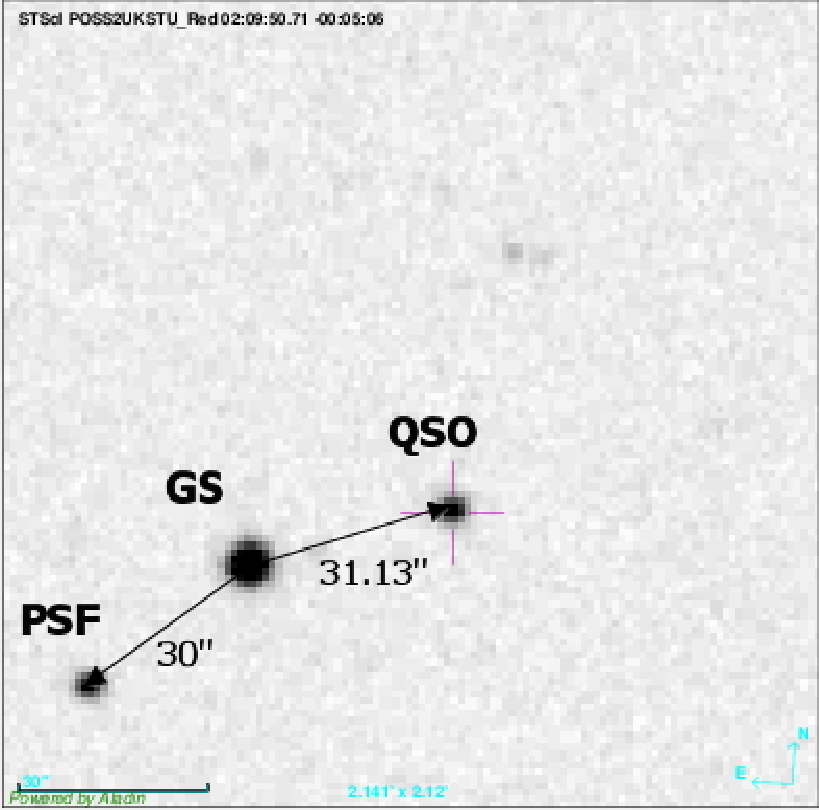}
   \includegraphics[width=59mm,height=58mm]{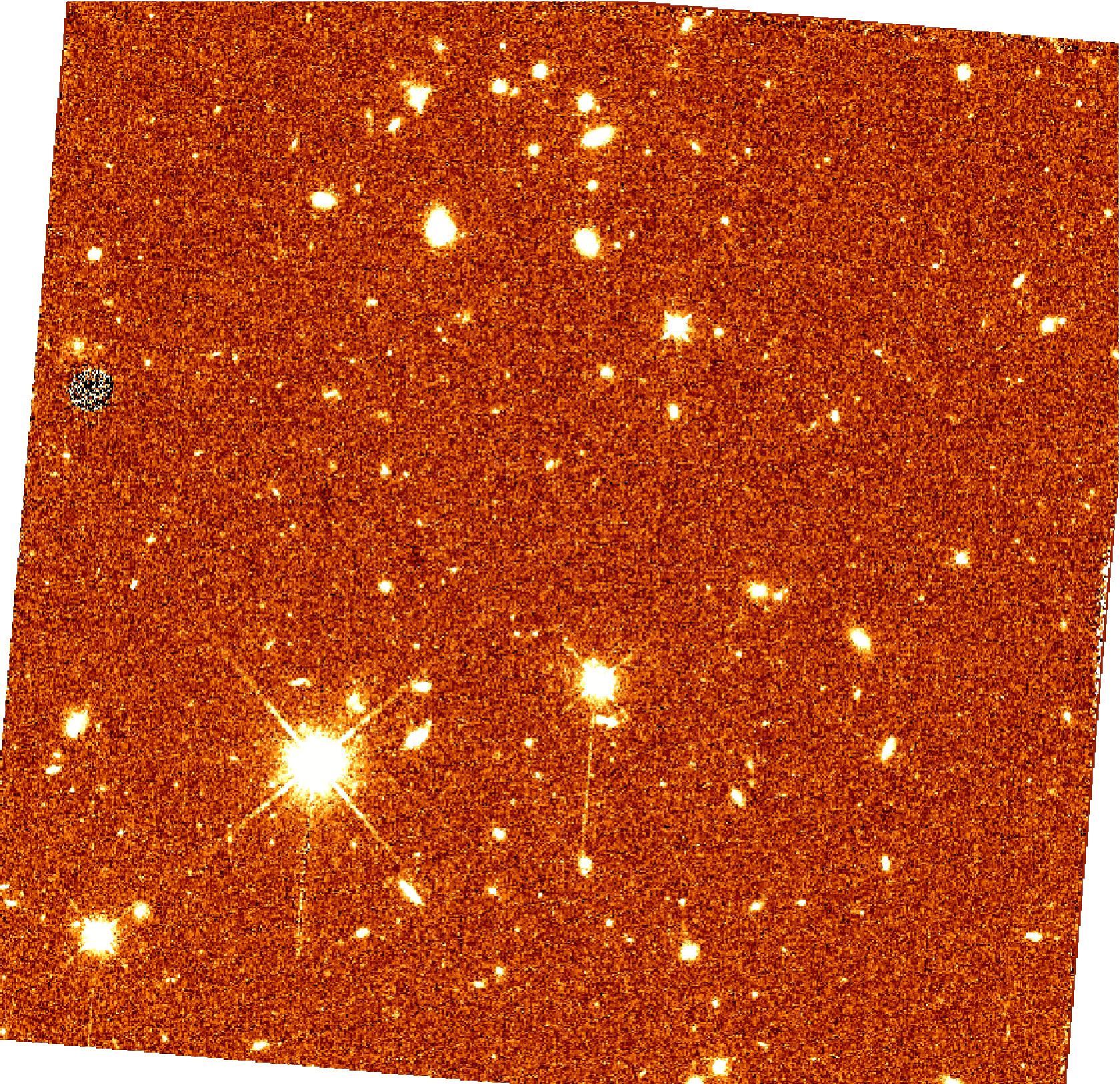}
  \end{center}
  \caption{Left: Finding chart of QSO, guide star (GS) and PSF calibration star from STScI POSS2UKSTU-Red; Right: A dither-combined WFC3/F140W image of
a region of $\sim 2'\times 2'$ around QSO UM402, from one orient of a single orbit.  The pixel scale is $0\secpoint 13$. North is up and East to the left.}
  \label{fig:1}
  \end{figure}

Usually, the guide star could not be used to directly model the PSF. This is because the PSF is expected to change with the angular distance from the AO guide star, and the actual PSF at the position of the target will be degraded. On the other hand, there will be the saturation problem for a very bright guide star required by the optimal AO correction. In this case, we have to select other PSF calibration star which could be observed at a condition as similar as possible to that of the QSO, i.e. similar magnitude, similar direction and angular distance to the guide star. We selected a suitable PSF calibration star for UM402, which is of similar brightness, similar guide star distance, but a guide star angle offset $\sim 180\degpp\,\,$ to that of QSO (see Fig. \ref{fig:1} left). We understand that the decrease in Strehl is not isotropic, and the shape of PSF varies over the whole field of view. However, Kamann (2008) suggested that it is possibly a good choice to select a calibration star around the mirror position of the quasar to overcome the variability of the PSF in large part, based on detailed studies on the residuals of AO-corrected images after subtraction of central PSF of the field of view. The details of the QSO, the guide star and the PSF calibration star are listed in Tab. \ref{tab.1}. The guide star used for AO correction is the same for QSO and the PSF calibration star.

\subsection{Observations and data reduction}
\label{sect:obs}
The AO-assisted ${\it Ks}$ band deep imaging of UM402 was made on
Sept.17-19, 2003 (UT), using the IRCS camera on
Subaru 8.2m telescope at Mauna Kea and the Subaru Cassegrain AO system with a 36 element curvature
wave front sensor, as well as a bimorph-type deformable mirror with the same number of elements (AO36; Takami et al. 2004).
The camera uses one $\rm
1024\,\times\,\,1024$ InSb Alladin III detector and has two imaging modes with different pixel scales. We adopted in the observation
a pixel scale of $0\secpoint 023$ (23mas mode),
providing a field size of $23\secpp \times \, 23\secpp$ (IRCS; Kobayashi et al. 2000).  In order to remove the bad pixels, we adopted nine-point dithering in a $3 \times 3$ grid with a dithering step of $5''$, $6\secpoint 5$ or $7''$. To reduce the readout noise, 16 times nondestructive readout (16-NDR) was applied for each readout. Dark frames and dome flats were taken at the end of each nights. Most of observing nights were clear and photometric. The median seeing size was $\sim 0\secpoint 5$, and the airmass was mostly smaller than $1.4$. After an optimal function of the AO system was achieved, we offset the FOV of the telescope to put the QSO or the PSF calibrator star in the center of the FOV. 

Similar as other currently available AO systems, we suffered from the small field view of the AO detectors, and were not able to include simultaneously a suitable PSF calibration star in the QSO exposures to directly evaluate the PSF. In order to monitor and assess the temporal variability of the PSF, we observed the PSF calibration star just before and after the observations of the QSO. More specifically, we observed the QSO itself in an exposure set of $9\times 80s$, or $9\times 70s$ using the $3 \times 3$ dithering pattern, nested between similar dithering observations of the PSF calibration star. Such interleaving observations could provide us information on the temporal variation of the PSF during the target observations. Meanwhile, such non-simultaneous PSF calibration would provide very similar correction quality of the AO systems to the QSO images, since the variability applies to both the PSF star exposures and the QSO images displaying similar Strehl values.

We adopt the ``core width $r_{20}$'', which is defined to encircle $20\%$ of the total flux of a point source, as the image quality indicator based on a close relation between $r_{20}$ and the Strehl ratio given by Kuhlbrodt et al. (2005). We measured the $r_{20}$ value of the QSO and the PSF calibration star for each exposure frame of all three observing nights, and adopted only good exposures with $r_{20}<3.5\,\,{\rm pix}$ (Strehl ratio $S\sim 50\%$ according to the empirical relation on $r_{20}$ vs. Strehl ratio). Our further analysis and discussion on the host galaxy properties mostly relies on these good exposures.

The package ``IRCS-IMGRED'' was used to make the dark frame, flat frame, bad pixel mask and sky frame, as well as flat fielding and sky subtraction (Minowa et al. 2008).
Finally, the dithered frames of the QSO and the PSF calibration star were aligned and averaged respectively, using an outlier rejection algorithm. The FWHM of the AO-corrected and combined QSO image by this run is $\sim 0\secpoint 13$, and $\sim 0\secpoint 11$ for calibration star images (see Tab. \ref{tab.1}).

In order to constrain the redshift range of the resolved faint galaxy of the QSO sightline, we obtained further deep J-imaging of the field centered on UM402 using IRCS camera and AO188 on Subaru telescope on Oct. 5, 2012 (UTC), in a service mode. We adopted in this run a pixel scale of $0\secpoint 052$(52mas mode), providing a field of view of $54\secpp \times \,54\secpp$. The weather condition during the observation was very good and the achieved FWHM in J-band is $\sim 0\secpoint 2$. Total exposure time for the target is two hours. We show the final combined J-band and Ks-band QSO images in Fig. \ref{fig:2} (upper).

To estimate the Strehl value of the AO-corrected images, we assume that PSF is approximated by a double 2d-gaussian profile as:$f_{psf}= \frac{f_{core}}{2\pi\sigma_{core}^2} exp[-(\frac{x^2}{2\sigma_{core}^2}+\frac{y^2}{2\sigma_{core}^2})]+\frac{f_{halo}}{2\pi\sigma_{halo}^2} exp[-(\frac{x^2}{2\sigma_{halo}^2}+\frac{y^2}{2\sigma_{halo}^2})]$, where $f_{core}$ is the flux ratio of diffraction core corrected by AO system to total flux; $f_{halo}$ is flux ratio of uncorrected (seeing limited) halo to total flux, and $f_{halo}+f_{core}=1$; $\sigma_{core}$ and $\sigma_{halo}$ are spatial broadening of AO-corrected core and uncorrected halo respectively. We fitted the AO-corrected and combined ${\it J \& Ks}$-band QSO or calibration star images using Markov Chains Monte Carlo. The Strehl ratio is estimated as: $Strehl=f_{core}+ f_{halo}\times\,(\frac{\sigma_{core}}{\sigma_{halo}})^2$, which gives a value of $\sim 39\%$ for the J-band image and $\sim 41\%$ for the Ks-band image.

The standard star FS110, p533-d and p338-c were observed as the photometric
calibrator, which were selected from Hawarden et al. (2001). 

\subsection{PSF construction and subtraction}
Our goal with the high resolution AO images is to detect faint galaxies along the QSO sightline, as well as the faint extended host galaxy hidden in the glare of the central bright QSO light. Therefore, it is mandatory to estimate properly the AO PSF and subtract the light contribution from the bright central point source of the QSO images, in order to unveil the underneath faint objects, and to reduce the effects to their photometric measurements.

For the deep Ks-band images, the PSF calibration star was observed as we have described in section \ref{sect:obs}. Thus, we can create a PSF using interleaved exposures of the PSF calibration star between QSO observations in two ways as following: 

1) Firstly, we constructed a sigma-clipping averaged PSF from the series of PSF calibration star images. To detect the host galaxy, we subtracted the PSF from the QSO image using a very conservative and simple method, same as other studies on the high-z QSO host galaxies. We scaled the PSF flux to the QSO central peak intensity and aligned them. Such a method implies an oversubtraction of the nuclear component from the inner region, and provides a model independent host detection and a low limit on the host flux (Sanchez et al. 2004). The PSF subtracted residual image of this bright QSO in the $Ks$-band was shown in Fig. \ref{fig:3} (right). The contour plots of the PSF subtracted QSO images for each observing night of Sept. 17 - Sept. 19 , as well as the coadded residual images of all three nights, were shown in Fig. \ref{fig:4} and Fig. \ref{fig:5} (left). 

2) Secondly, we applied a principle component analysis based on Karhunen-Lo$\grave{e}$ve (KL) transform to construct a PSF from the series of interleaving exposures of the calibration star (Karhunen 1947, Lo$\grave{e}$ve 1948). This algorithm is adopted to quantify the modes in which the PSF varies with time by a basis function that characterizes the temporal variations of the calibration star as well as the QSO images (Chen et al.2006, Soummer et al. 2012). A $\sim 2\secpoint 5 \times 2\secpoint 5$ region centered on each reduced frame of the calibration star image was selected as a search area, which is considered free from any other astronomical signal. We computered the basis function by calculating the KL transform of the set of calibration PSFs over the ensemble of searching areas. The first 20 modes with the largest eigenvalues of the basis function were selected to construct a best estimation of the actual PSF of the QSO image from the projection of the QSO image on the KL basis. We understand that the light from the QSO host galaxy or absorbing galaxies might be mistakenly interpreted as components of the PSF, and thus be oversubtracted from the QSO image. Detailed simulation and discussion on this point will be presented by our next paper (He et al. in preparation).  In this work, we simply scale the QSO and the PSF model to make their peak fluxes to be equal. In this way, it is consistent with the analysis using the first approach, as well as previous studies on high-z QSO host galaxies. The results of this analysis were shown in Fig. \ref{fig:4} and Fig. \ref{fig:5} (right). We noticed that the residuals of the PSF subtracted images are mostly similar in the morphological structure for the two different approaches of PSF reconstruction, as well as among three consecutive observing nights, although slightly shrinked in case of principle component analysis. We thus consider that the host galaxy of this bright radio-quiet QSO towards the peak epoch of galaxy formation at $z\sim 3$ has been marginally resolved. 

In the interests of obtaining only $J-Ks$ color of the resolved objects along the QSO sightline, the PSF calibration star was not observed exclusively for the deep J-band imaging using IRCS+AO188, during the service run on Oct.5, 2012(UTC). Given that IRCS J-band imaging is looking at the wavelength blueward of the 4000\rAA Balmer break, we assume that the extended emission from the host galaxy of this QSO is negligible within 2hr exposures. In this case, a PSF could be constructed from the series of QSO images after carefully masking out the resolved faint galaxies in the QSO field, using a principle component analysis. We finally subtracted the model PSF from the coadded QSO image in a way similar as we have done for the Ks-band images. The PSF subtracted QSO image in the J-band is shown in Fig.3 (left).

\begin{table*}
\caption{Observed QSO, PSF star and the guide star}\label{tab.1}
\begin{tabular}{lcccclcccc}
{Obj.} & {Type} & {RA(J2000)} & {DEC(J2000)} & {z} & {$R_{mag}$} &
{$ t_{exp} $(hrs)$^{a}$}  & {FWHM$^{b}$} & {$K_{s}^{c}$} & {GS(d)$^{d}$}\\
\hline
~UM 402         & RQQ & 02 09 50.71 & -00 05 06.6 & 2.855   & $15.8$  & 3.0 & 0.13 & 14.54 & 31 \\
~PSF star       &     & 02 09 54.51 & -00 05 34.0 &         & $16.6$  & 1.0 & 0.11 & 15.47 & 30 \\
~Guide star     &     & 02 09 52.84 & -00 05 15.2 &         & $13.8$  &      &      &       &    \\
\hline
\end{tabular}
\\
$^a$ Good exposure times\\
$^b$ Image quality measured as the FWHM for the coadded ${\it Ks}$ images of all good exposures in arcseconds.\\
$^c$ Observed  K magnitude of the target for the coadded Ks images. \\
$^d$ Distance in arcsec from the QSO (PSF star) to the Guide star.
\end{table*}

\subsection{Archived WFC3/F140W images}
WFC3/F140W imaging of this QSO field with the HST is available from the MAST archive (PI: Dawn Erb, HST proposal ID 12471). The IR wide F140W filter ($1.2\mu m -1.6\mu m$) covers the gap between J and H band which is inaccessible from the ground. The pixel scale of WFC3 images is $0\secpoint 13$/pixel. We retrieved the calibrated, flat-fielded four exposures from the archive (2 orients of a single orbit, 2 dithers for each orient), where the exposure time is 202.934 seconds for each exposure. Thus, total exposure time is 811.736 seconds for the WFC/F140W images.

We reduced the WFC3/F140W images in two situations: 1) combined directly the 4 frames after de-rotating the two dithering frames of the second orient by 30 degrees and aligning all the four frames (see Fig. \ref{fig:2} bottom); 2) performed PSF subtraction to the QSO images in a classical method presented by Rajan et al. 2011. Specifically, we generated one QSO image by combining the aligned two dithers of each orient. A PSF image was constructed by median combine of aligned frames of other orients, and subtracted from the QSO image. The PSF subtracted QSO images were combined into a final image after de-rotating the second orient by 30 degrees. We show the processed coadded images in Fig. \ref{fig:3} (bottom). 

\section{Analysis}
\label{sect:analysis}
\subsection{Galaxies near the QSO sightline and the luminosity}
\begin{figure}[h]
  \begin{center}
   \includegraphics[width=59mm,height=58mm]{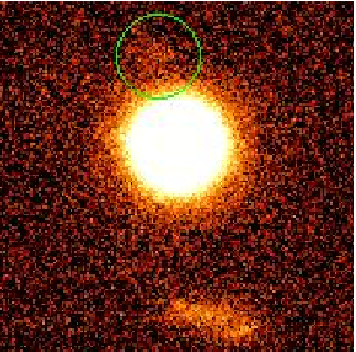}
   \includegraphics[width=59mm,height=58mm]{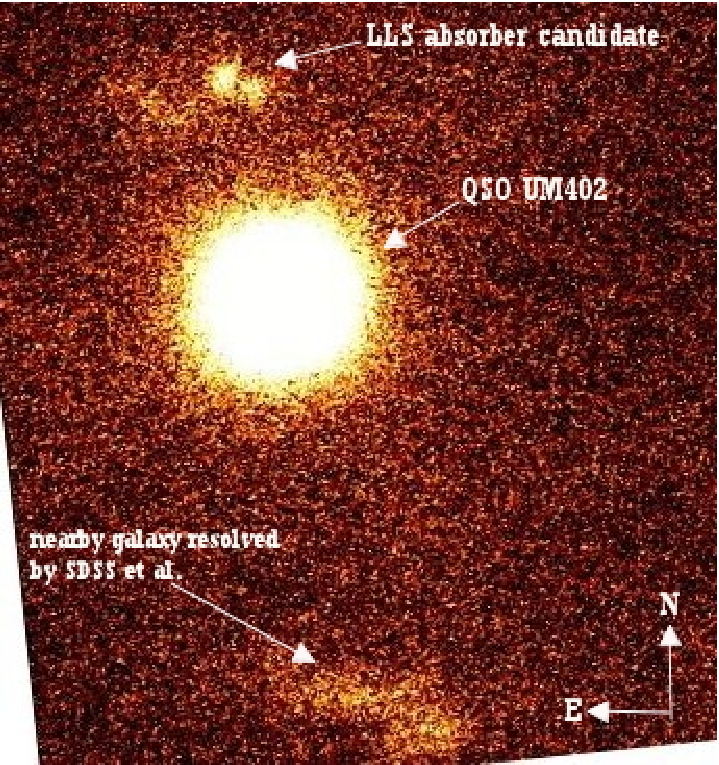}
   \includegraphics[width=59mm,height=58mm]{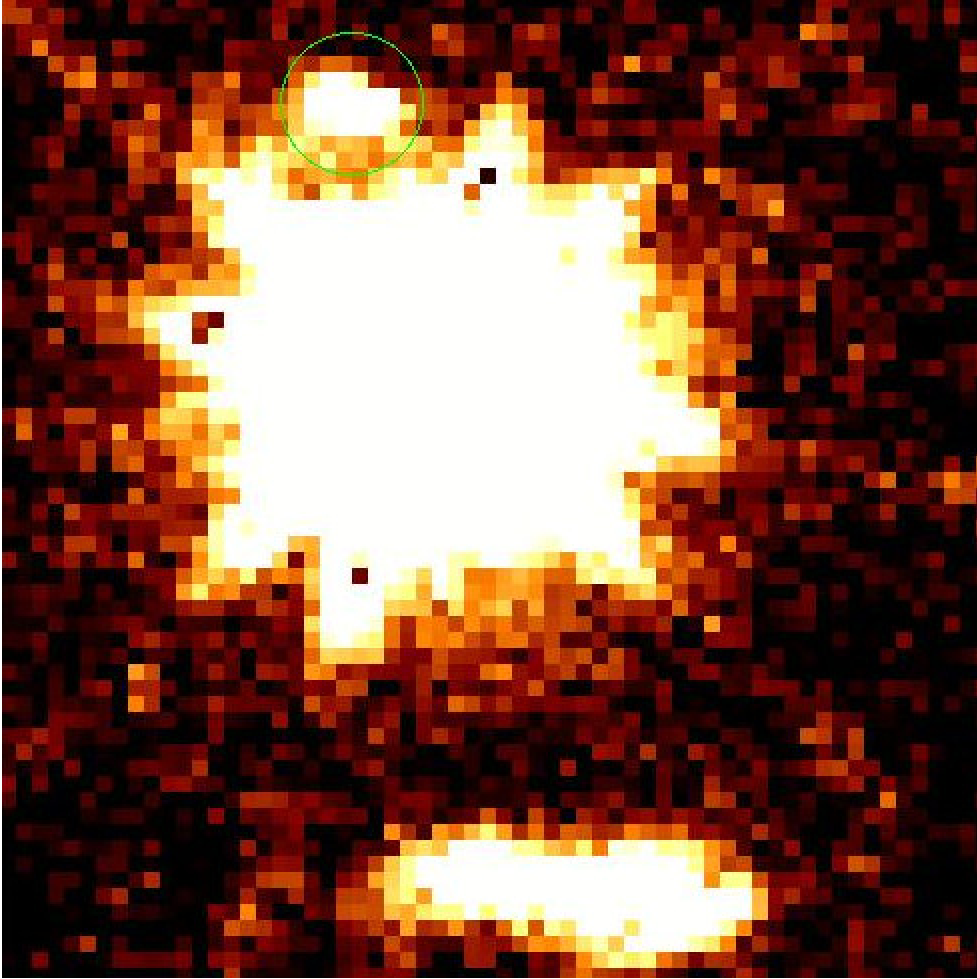}
  \end{center}
  \caption{Upper left: A combined image of a region of $\sim 5''\times 5''$ around QSO UM402 (central bright source) in the ${\it J}$-band. The LLS absorber candidate is $\sim 2\secpoint 4$ north of the QSO sightline (marked by a circle in the figure). The pixel scale of the J-band image is $0\secpoint 052$ and the AO-corrected FWHM $\sim 0\secpoint 2$; Upper right: A combined ${\it Ks}$-band image of a similar region centered on QSO as that of the ${\it J}$-band image shown in the left. The LLS absorber candidate is indicated in the figure, and appear as a merging system with two close components of separation of $\sim 0\secpoint 3$. The galaxy southern of the QSO sightline has been detected by SDSS and other optical deep imaging as a nearby irregular galaxy at $z\sim 0.36$ (Le Brun et al. 1993, Guillenmin \& Bergeron 1997). The pixel scale of ${\it Ks}$ image is $0\secpoint 023$, and the AO-corrected FWHM $\sim 0\secpoint 13$; Bottom: A combined WFC3/F140W image of UM402 (a pixel scale of $0\secpoint 13$), after aligning and de-rotating the two dithered frames of second orient. The LLS candidate galaxy is indicated in the figure by a circle . North is up and East to the left.}
  \label{fig:2}
  \end{figure}

The high resolution images in the $J\& Ks$ band, as well as WFC3/F140W filter, have clearly resolved two galaxies within a $\sim 5\secpp \times 5\secpp$ field around UM402 (see Fig. \ref{fig:2}). They are a fuzzy galaxy about $4\secpoint 7$ southern of the QSO, and a close object $\sim 2\secpoint 4$ northern of the QSO sightline. In the deep $Ks$-band images (Fig. \ref{fig:2} upper right), the faint object which lies $\sim 2\secpoint 4$ northern of the QSO sightline, appears as a double system with a separation of the two components $\sim 0\secpoint 3$. Meanwhile, there seems to be a faint tidal-tail like feature towards southeast for the left component of this double system, suggesting a possible merging system for this object. The fuzz galaxy southern of the QSO sightline has been detected by SDSS and other optical deep imaging as a nearby irregular galaxy at $z\sim 0.36$ (Le Brun et al. 1993, Guillenmin \& Bergeron 1997).

\begin{figure}[h]
  \begin{center}
   \includegraphics[width=59mm,height=58mm]{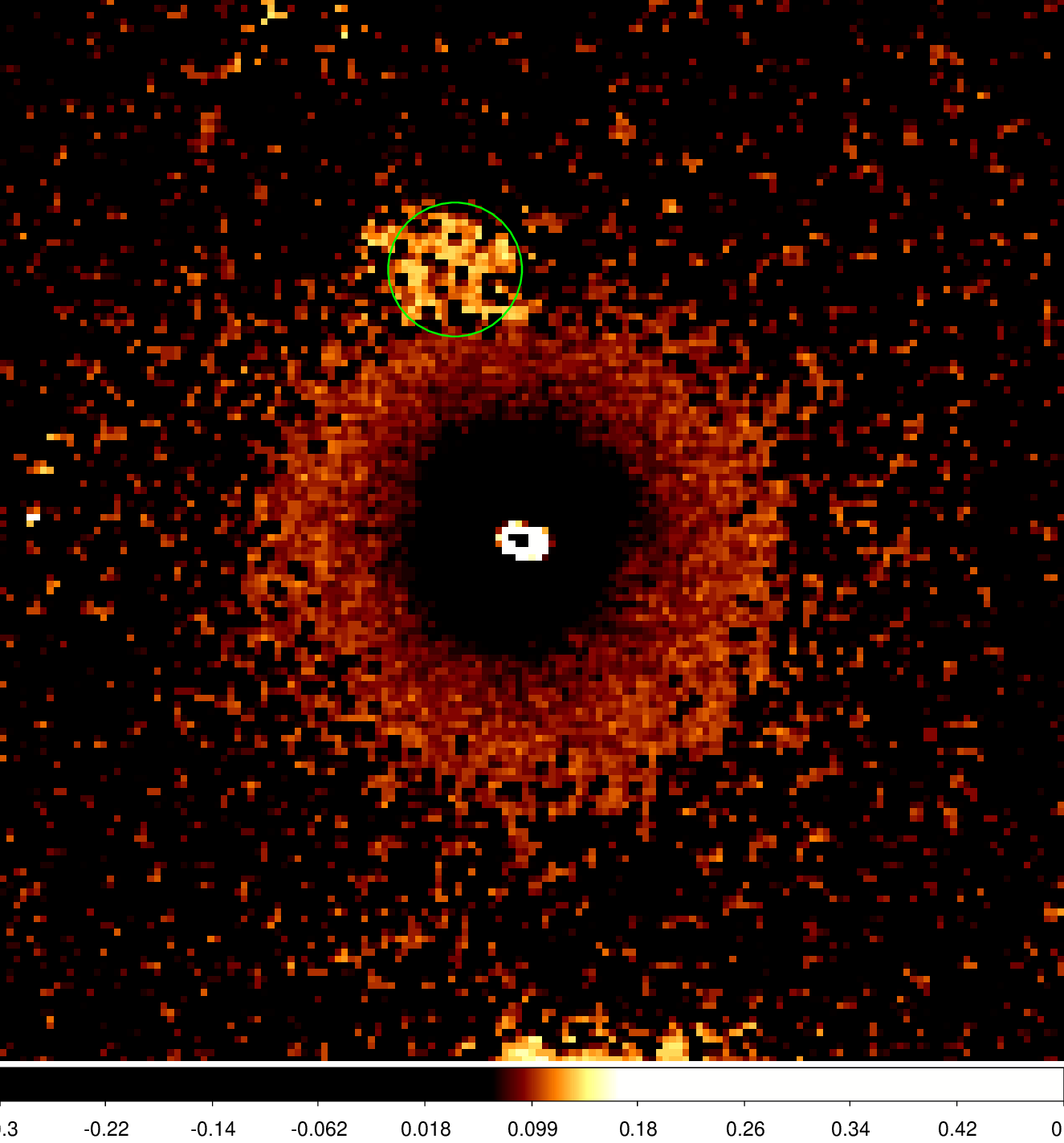}
   \includegraphics[width=59mm,height=58mm]{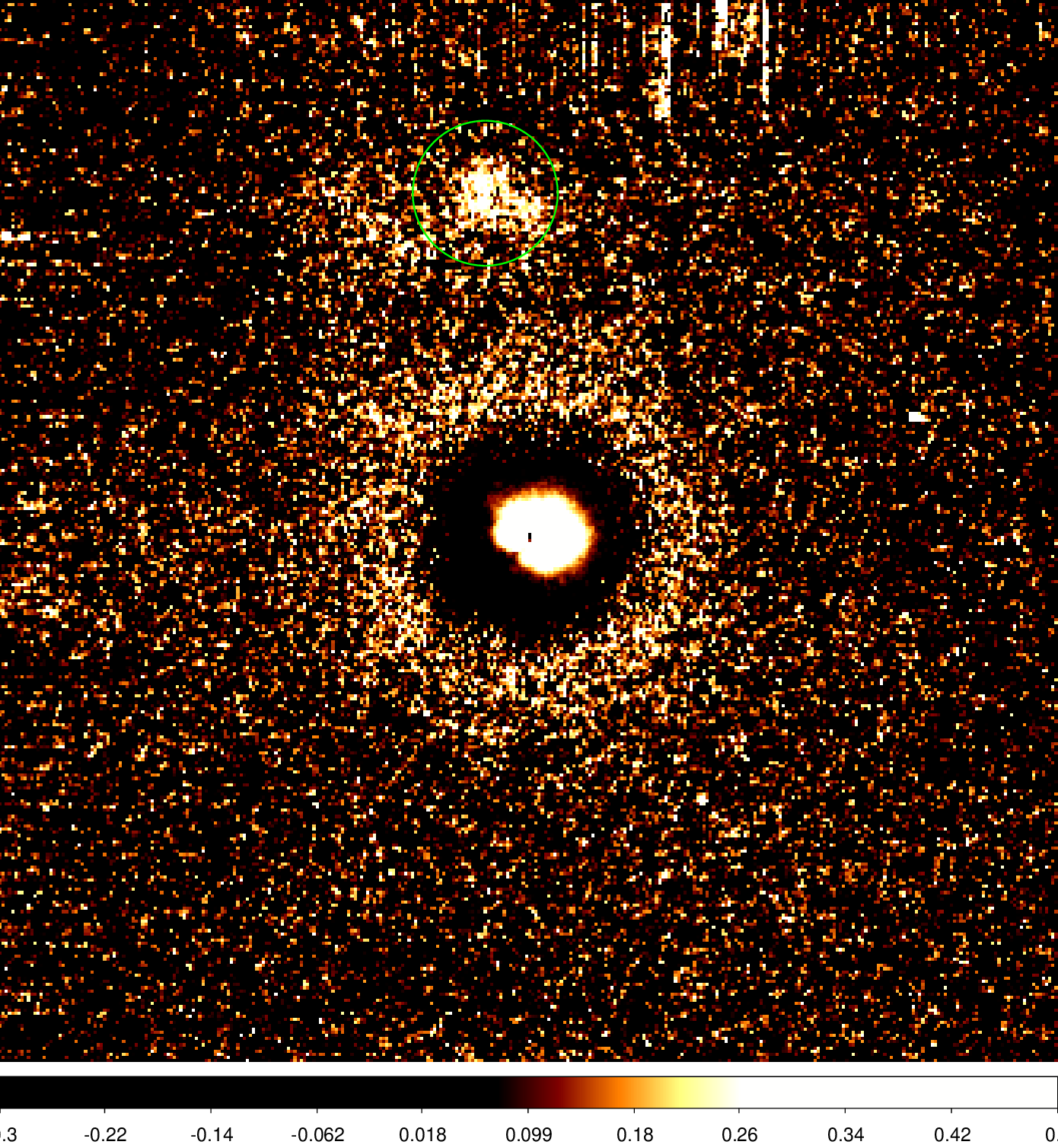}
   \includegraphics[width=59mm,height=58mm]{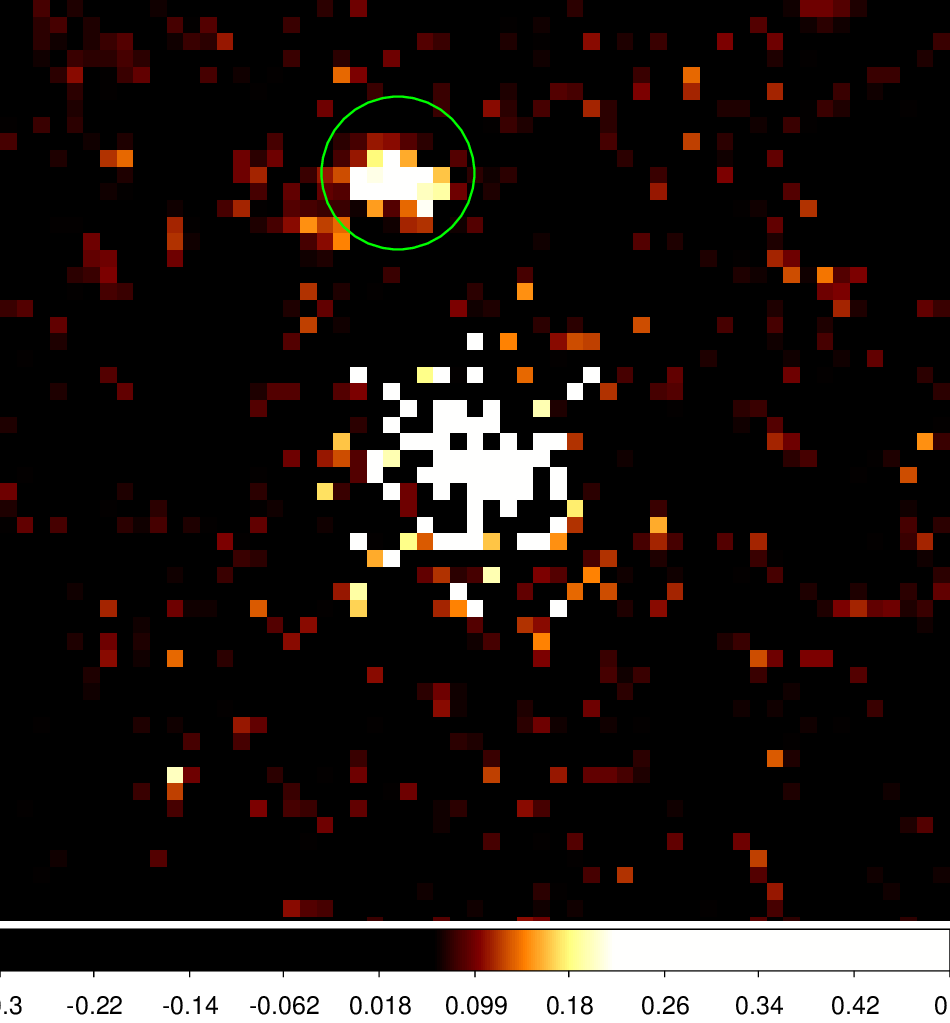}
  \end{center}
  \caption{Upper left: A PSF subtracted and coadded J-band image ($0\secpoint 052/pix$) around QSO UM402. The construction of PSF model was described in the last paragraph of section 2.3; Upper right: A PSF subtracted and coadded Ks-band image of a similar region centered on QSO as that of the J-band image shown in the left ($0\secpoint 023/pix$). We constructed an averaged PSF from the series of PSF star images as described in section 2.3 (the first approach); Bottom: A PSF subtracted and coadded WFC3/F140W image of UM402 ($0\secpoint 13/pix$), after aligning and de-rotating the PSF subtracted images. The PSF was constructed in a classical way by median combine of aligned frames of other orients (Rajan et al. 2011). The LLS candidate galaxy is indicated in the figure by a circle. The field of view of the image is $\sim 7\secpoint 5 \times \, 7\secpoint 5$. }
  \label{fig:3}
  \end{figure}

Using SExtractor, we measured the photometry of this faint object in the coadded and PSF subtracted images of all three datasets shown in Fig.\ref{fig:3}, with a detection threshold of $2.5\sigma$ over the sky level for the $Ks$-band image, a detection threshold of $2\sigma$ for the WFC3/F140W image , as well as a $1.5\sigma$ detection threshold for the deep J-band imaging. The photometric results of this faint galaxy measured with a small diameter aperture ($0\secpoint 6$) and a large diameter aperture ($1\secpoint 2$) are presented in Tab. 2. However, we noticed that the accuracy on PSF subtraction would cause systematics to the photometry of the faint objects along the QSO sightline.

For a more reliable photometric measurement against neighboring contamination in the QSO nearby field, we also performed a 2-D decomposition algorithm GALFIT on the coadded $Ks$-band image, where the QSO(moffat), and any other nearby galaxies (S$\acute{e}$rsic profiles) in the field were fitted simultaneously to deblend everything together, in order to reduce the contaminating flux from the wing in the PSF of the QSO (Peng et al. 2002). The best fitting gives an apparent magnitude $m_{k}=23.32\pm 0.05$ for the faint galaxy which lies $\sim 2\secpoint 4$ northern of the QSO sightline, mostly consistent with the measurements using SExtractor which we discussed above. 

To estimate the color of this faint object, we adopted coadded and PSF subtracted ${\it Ks}$-band image as the detection image and a detection threshold of $\mu =2.5\sigma$ of the skylevel (Bertin \& Arnouts 1996).  The ${\it J-Ks}$ color were determined by re-running SExtractor in the double-image mode, in which the faint object detected on the ``detection image''
(in the ${\it Ks}$-band) was measured with the same aperture in the registered J-band image, giving a result of ${\it (J-Ks)_{vega} \sim 1.6}$.

According to the impact parameter vs column density relation ($b\,\,-\,\,logN_{HI}$) for all confirmed DLA and LLS absorbers given by Moller \& Wallen (1998), we suspect this close double system might be a candidate galaxy giving rise to the Lyman Limit absorption at $z_{abs}\sim 2.5$ previously seen in the QSO spectrum ($N_{HI}>4.6\times10^{17}cm^{2}$). Considering that the apparent K-band magnitude vs. stellar mass relation for objects at $2.3<z<2.6$ from the MOIRCS Deep Survey (MODS), this galaxy would have a stellar mass of $\sim$ $3\times 10^{9}\,M_{\odot}$, at the low-mass end of the MODS sample (Tanaka et al. 2011).

Further observations for the spectroscopic redshift as well as kinematics of both components of this faint object are strongly required. If confirmed, it would be an important high-z evidence of a merging system as Lyman Limit absorber.

\begin{table*}
\caption{Photometric results of the LLS absorber candidate}\label{tab.2}
\begin{tabular}{lcccccc}
{aper. diameter} & {J app. mag.} & {J abs. mag.} & {F140W app. mag.} & {F140W abs. mag.} & {Ks app. mag.} & {Ks abs. mag.}\\
\hline
~$0\secpoint 6$ & $26.40\pm 0.14$ & -20.21 & $24.01\pm 0.37$ & -22.60 & $24.25\pm 0.07$ &  -22.00  \\
~$1\secpoint 2$ & $25.43\pm 0.10$ & -21.18 & $23.65\pm 0.32$ & -22.96 & $23.52\pm 0.05$ &  -22.63  \\
\hline
\end{tabular}
\end{table*}

\subsection{Extended emission from the host galaxy}

To detect the host galaxy, we carefully subtracted the PSF from the QSO image using two different approaches which we have discussed in Section 2.3. Since we simply and conservatively scaled the PSF peak flux to match the QSO central peak intensity and aligned them in both two approaches, the residuals after PSF subtraction from the QSO images implies an oversubtracted nuclear component of the inner region, which is a model independent host detection and a low limit on the host flux (Sanchez et al. 2004). 

\begin{landscape}
\begin{figure}\centering
   \includegraphics[width=75mm,height=62mm]{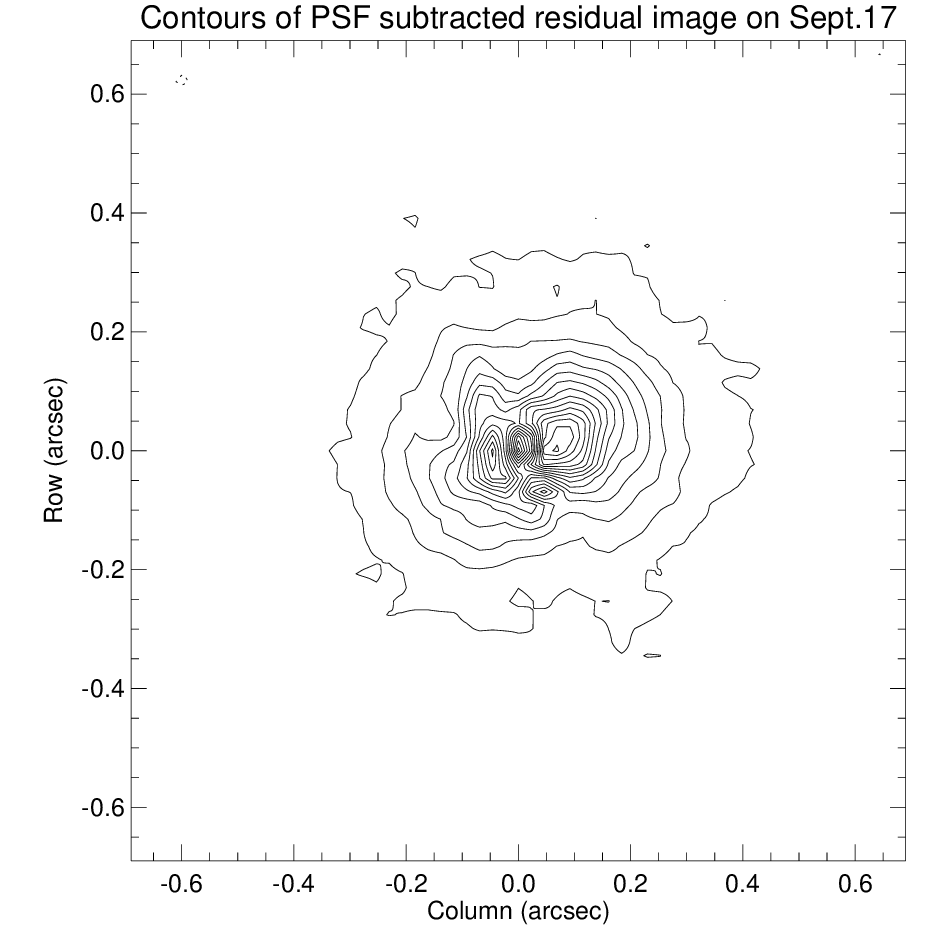}
   \includegraphics[width=65mm,height=64mm]{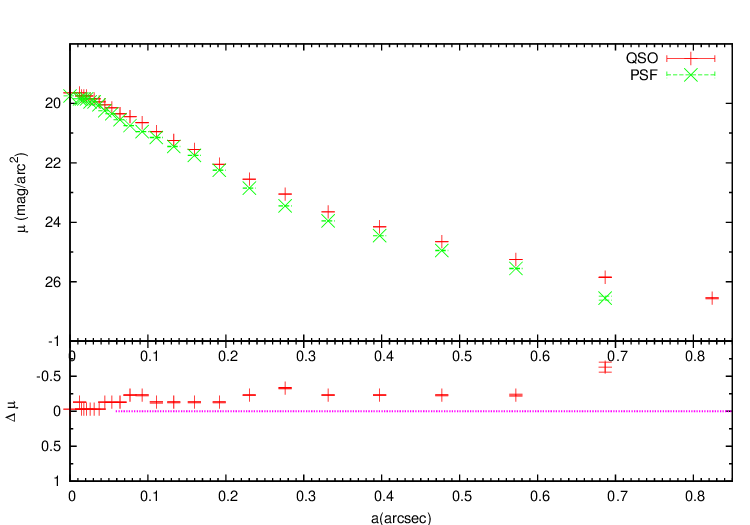}
   \includegraphics[width=65mm,height=64mm]{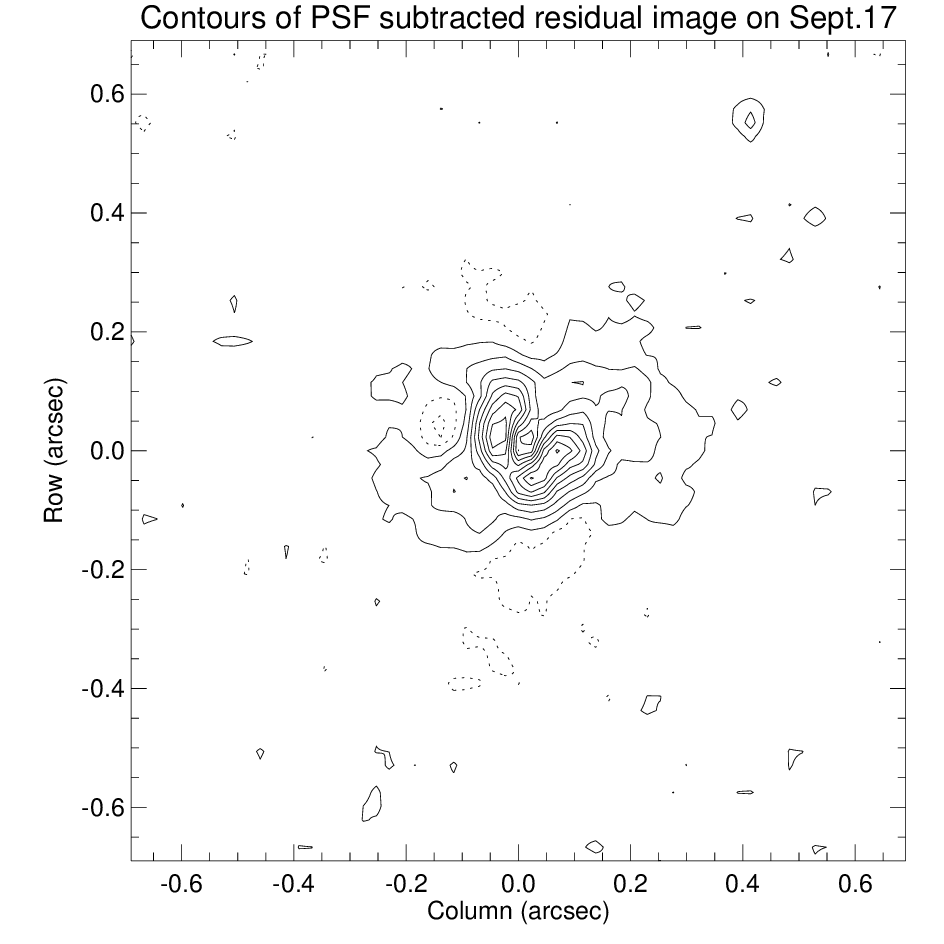}
   \includegraphics[width=75mm,height=62mm]{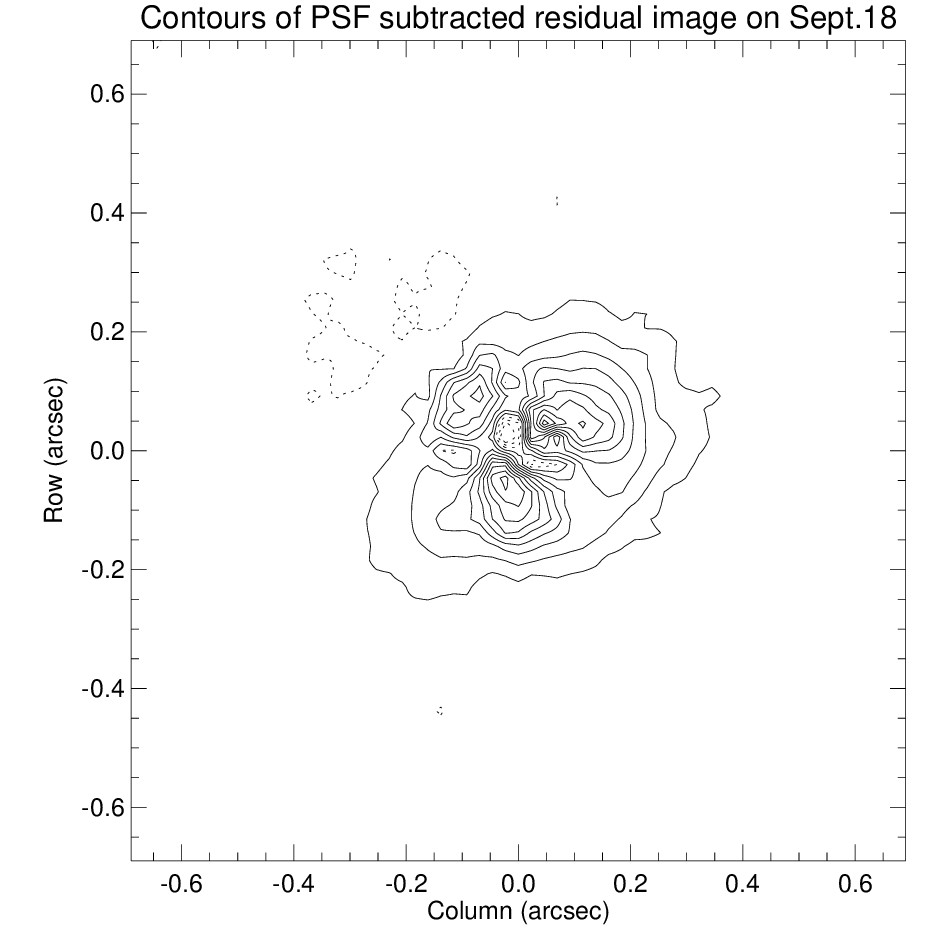}
   \includegraphics[width=65mm,height=64mm]{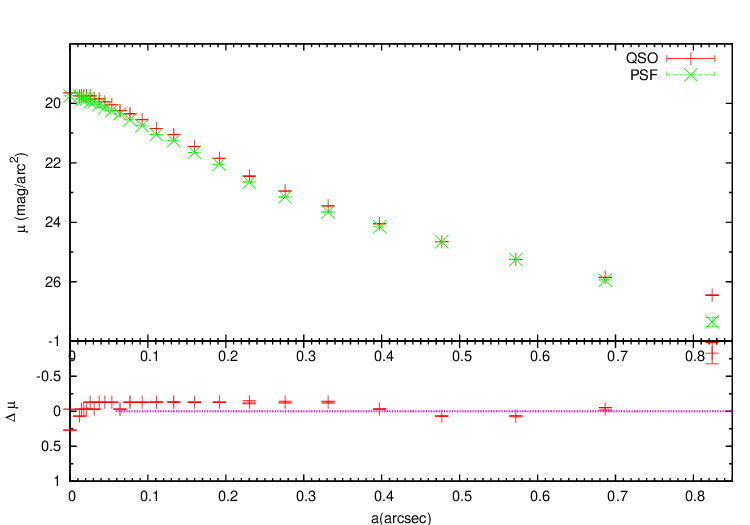}
   \includegraphics[width=65mm,height=64mm]{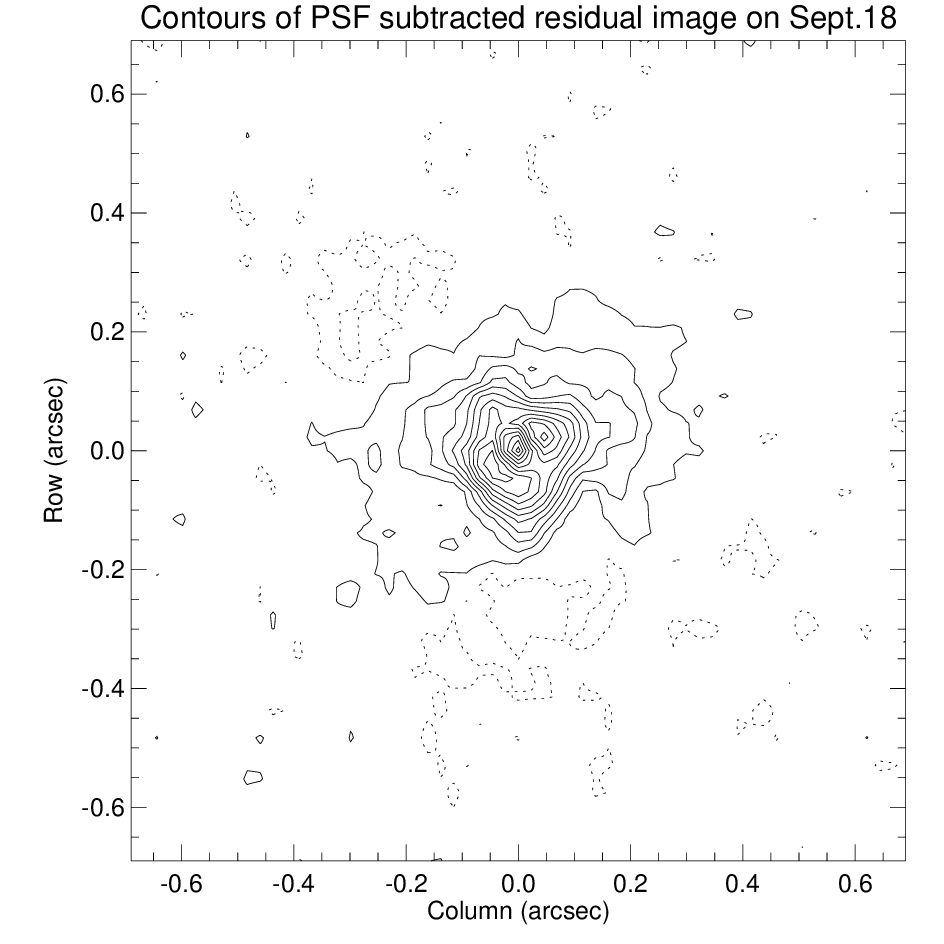}
  \caption{Left: From top to bottom, we show the plots of the extended emission of the quasar host revealed after subtracting a PSF from the combined ${\it Ks}$-band images taken on Sept. 17 --19 respectively, and that of all three nights. The PSF was contructed by averaging the PSF star images using an outlier rejection algorithm; Middle: The observed radial surface brightness profiles of the QSO (red plus) and the PSF star (green cross) are shown from top to bottom, for the coadded good images of each observing night from Sept. 17 to Sept. 19, as well as that of all three observing nights; Right: From top to bottom, we show the plots of the extended emission from the PSF subtracted good QSO images using approach 2) in section 2.3 for each observing night from Sept. 17 - Sept. 19, as well as that of all three nights. The PSF was constructed from the series of PSF star images using principle component analysis. The contour levels are from pixel value of -0.5 to 3.5 in counts, with an interval of 0.2. We plot out the negative contours in dashed lines.}
\label{fig:4}
\end{figure}
\end{landscape}

\begin{landscape}
\begin{figure}\centering
   \includegraphics[width=75mm,height=62mm]{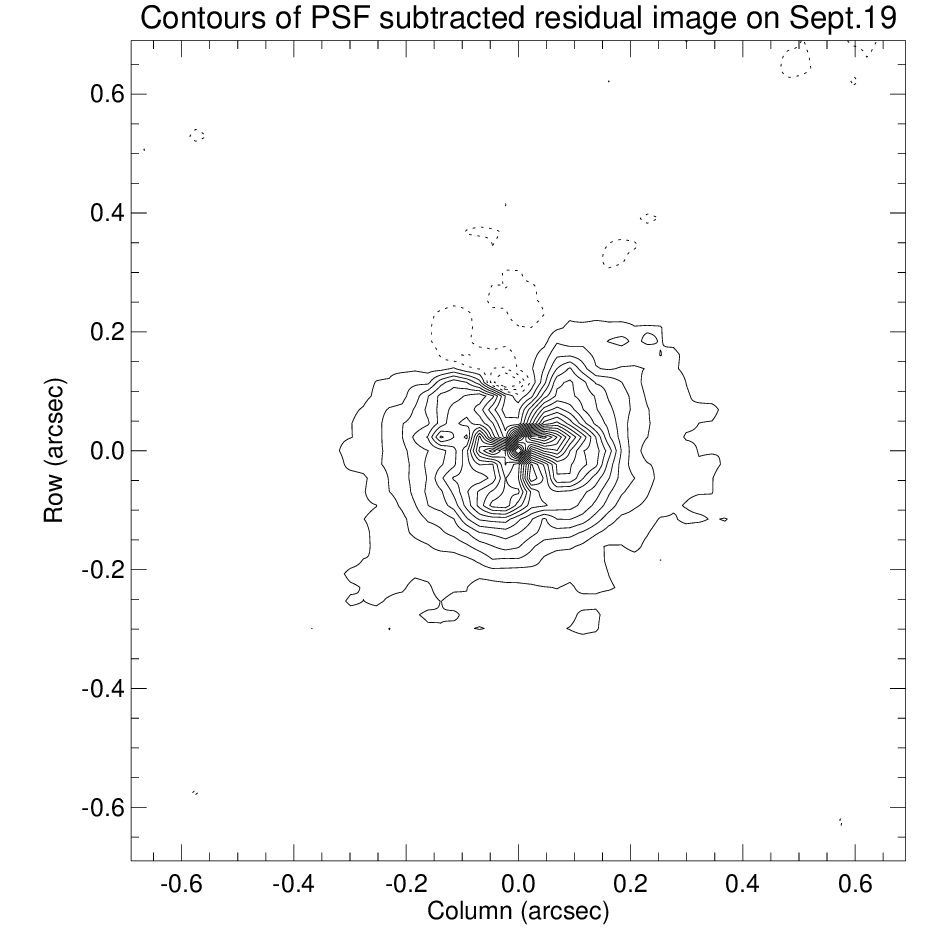}
   \includegraphics[width=65mm,height=64mm]{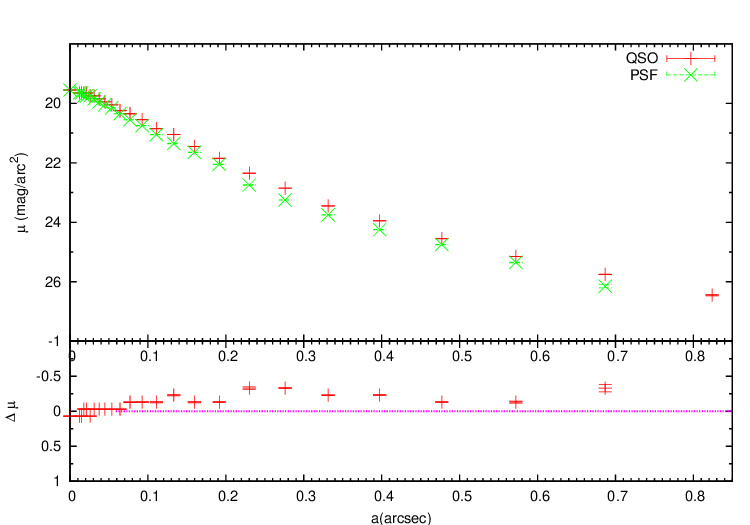}
   \includegraphics[width=65mm,height=64mm]{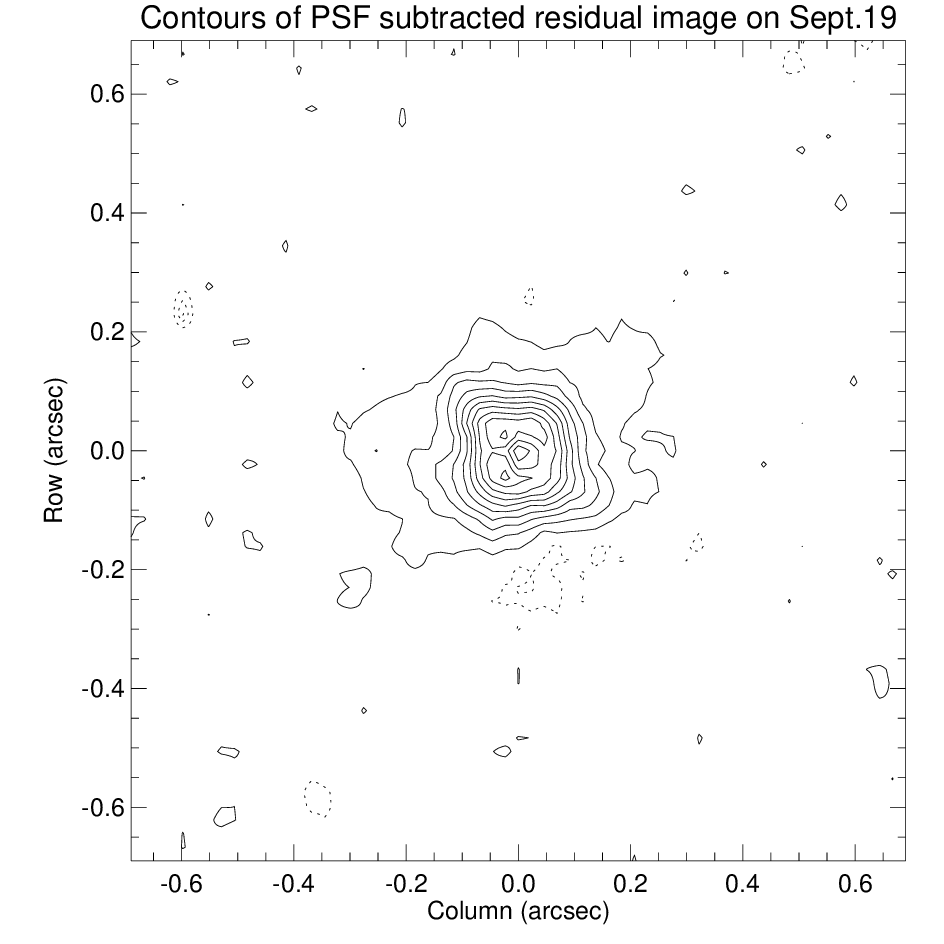}
   \includegraphics[width=75mm,height=62mm]{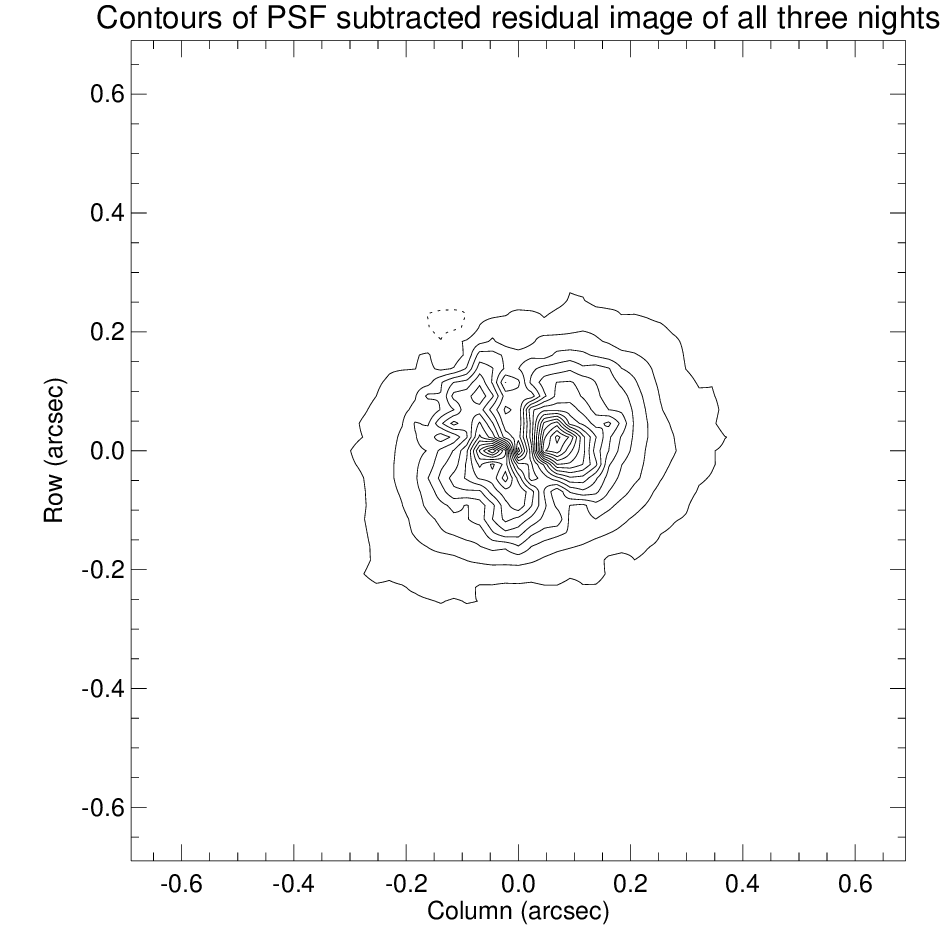}
   \includegraphics[width=65mm,height=64mm]{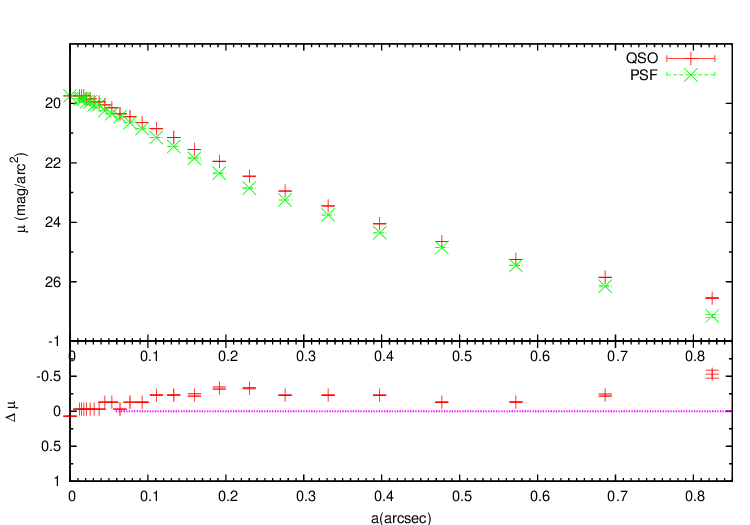}
   \includegraphics[width=65mm,height=64mm]{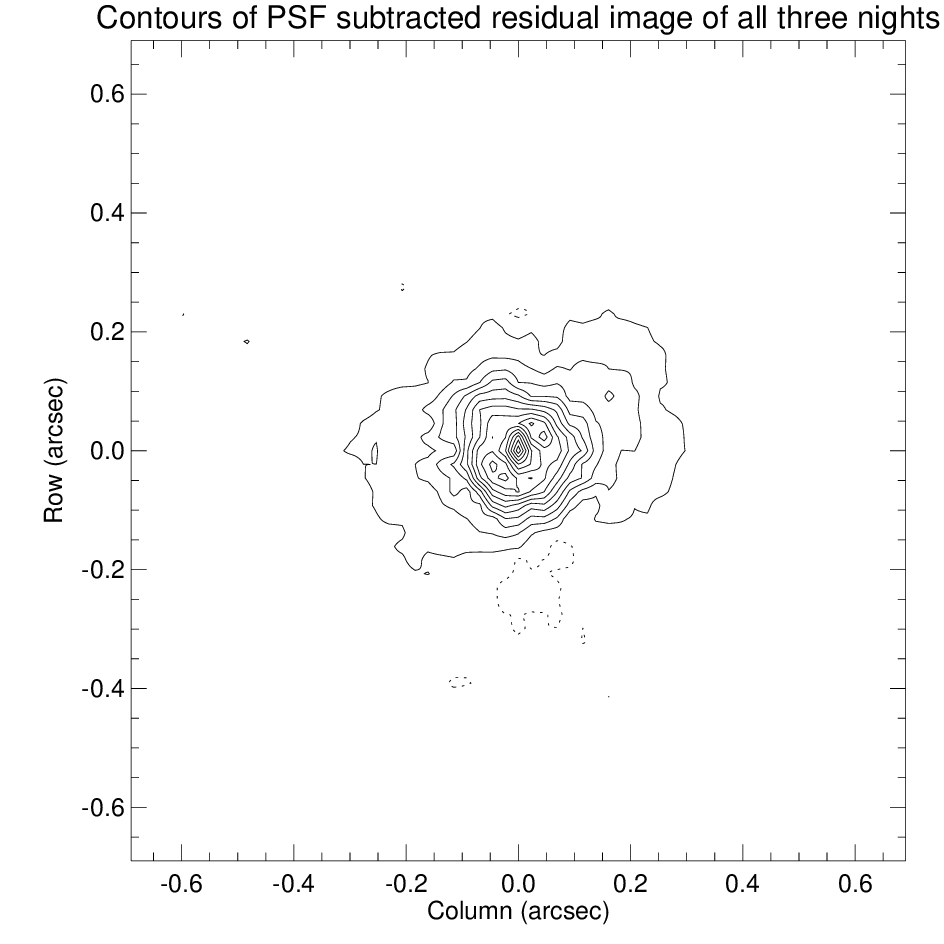}
  \caption{Fig. \ref{fig:4} continued}
\label{fig:5}
\end{figure}
\end{landscape}

The contour plots of the extended emission from the PSF subtracted co-added QSO images in the ${\it Ks}$-band using the first approach described in section 2.3 for each observing night from Sept. 17 - Sept. 19, as well as that of the PSF subtracted coadded residual images of all three nights were given in Fig. \ref{fig:4} (left) and Fig. \ref{fig:5} (left). A low limit of the host magnitude was estimated to be $m_{k}=22.3$, $m_{k}=23.2$, $m_{k}=22.5$ and $m_{k}=22.4$ for the images taken on Sept. 17-19 respectively, as well as that of all three observing nights, by photometry on the PSF subtracted QSO image using SExtractor package (a diameter aperture of $1\secpoint 2$). The systematic errors on the photometry of the host galaxy for different observing nights from Sept.17-19 reach about one magnitude. This indicates that PSF variation of different observing nights significantly affects the host detection and its photometric measurement. On the other hand, more reliable PSF reconstruction algorithm is required in this case.

In Fig. \ref{fig:4} (right) and Fig. \ref{fig:5} (right), we showed the contour plots of the extended emission from the PSF subtracted coadded QSO images using the second approach introduced in section 2.3 for each observing night from Sept. 17 - Sept. 19, as well as that of all three nights.  Here, a model PSF was constructed using principle component analysis based on Karhunen-Lo$\grave{e}$ve (KL) transform, and was subtracted from the coadded ${\it Ks}$-band QSO images in a consistent way as that of the first approach discussed in section 2.3, in order to double-check the results.  A low limit of the host magnitude was estimated to be $m_{k}=23.6$, $m_{k}=23.3$, $m_{k}=23.4$, and $m_{k}=23.3$ for the images taken on Sept. 17-19 respectively, as well as that of all three observing nights, by photometry on the PSF subtracted QSO image using SExtractor package (a diameter aperture of $1\secpoint 2$). We noticed that the measured host magnitude of the three consecutive observing nights are basically consistent, and the systematic errors on the photometric measurement of the host galaxy using a PSF reconstruction algorithm based on the principle component analysis (PCA) are reduced by more than a half. This is probably because the PSF reconstruced using PCA would account for most of PSF temporal variation, which might be smeared out in case of an averagely combined PSF.

We measured the radial profiles of the coadded images of the QSO and the PSF star ($r_{20}<3.5\,\,{\rm pix}$) using STSDAS task ELLIPSE, after masking out the close companions. The radial profiles of the coadded QSO images of each observing night are compared with the PSF stellar profiles acquired during QSO exposures of that night, and that of the coadded QSO and PSF exposures of all three nights, as well as their residuals are presented in Fig. \ref{fig:4} and Fig. \ref{fig:5} (middle). 

We noticed from Fig. \ref{fig:4} and Fig. \ref{fig:5} that the extended emission of the QSO host galaxy at the radii $> 0\secpoint2$ is consistently resolved for almost all three nights using three different analyses, indicating a marginally resolved host galaxy of this bright radio quiet quasar towards the peak epoch of QSO activity. We can accept roughly a low limit of host magnitude to be $\sim 23.3$ based on the analyses which we have discussed.

Finally, we zoomed out the contour plot on the left bottom of Fig. 5, to provide a larger size of the PSF subtracted residual image with the faint LLS candidate galaxy visible in the field of view. The LLS candidate galaxy is seen at the bottom of the plot, $\sim 2\secpoint 4$ below the extended emission of the quasar host.

\begin{figure}[h]
  \begin{center}
    \includegraphics[width=118mm,height=110mm]{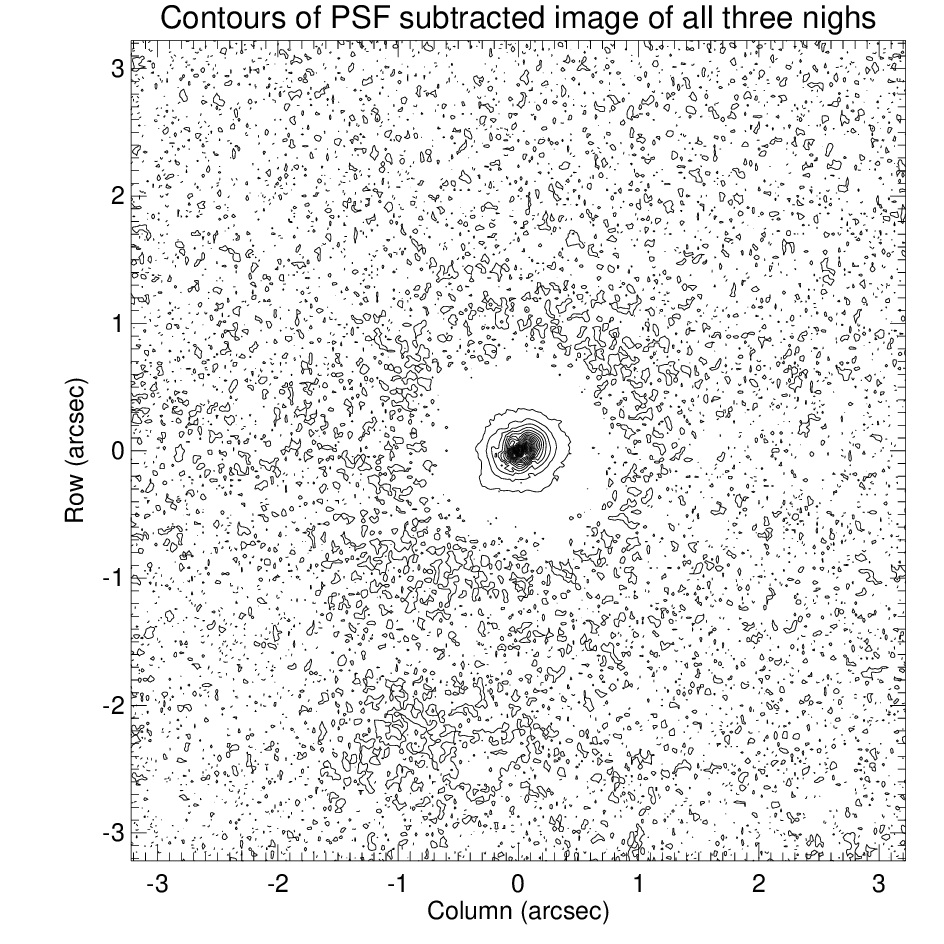}
  \end{center}
  \caption{The contour plot of a region of $\sim 6\secpp \times 6\secpp$ centered on QSO UM402 in ${\it Ks}$-band, after subtracting a PSF from the co-added images of all three observing nights. The PSF was contructed by coadding the series of nested exposures of PSF star using an outlier rejection algorithm (the first approach of section 2.3). The extended emission of the quasar host are marginally revealed, and the LLS candidate galaxy is visible at the bottom of the plot ($\sim 2\secpoint 4$ below the extened emission of the quasar host). The orientation, as well as the contour levels and interval of the plot are the same as that of Fig.4 and Fig.5. South is up, and East to the left.}
  \label{fig:6}
  \end{figure}

\section{Summary}
We have presented analyses of adaptive optics deep images in the ${\it J \& Ks}$ bands centered on QSO UM402 at $z_{em}=2.856$ using IRCS camera and the AO systems on Subaru telescope, as well as the WFC3/F140W archived images.

A faint galaxy ($\sim 2\secpoint 4$ north of the QSO sightline) has been clearly resolved by all three high resolution datasets. Especially for the deep ${\it Ks}$-imaging, it appears as a double system with a separation of the two components $\sim 0\secpoint 3$, while the faint tidal-tail like feature from the left component of the double sytem indicates that it is probably a merging system. According to the empirical relation on the impact parameter vs. neutral hydrogen column density of all confirmed DLA and LLS absorbers given by Moller \& Wallen (1998), as well as its red color $\rm (J-Ks)_{vega} \sim 1.6$, we suspect that this faint object might be a candidate galaxy giving rise to the Lyman Limit absorption at $z_{abs}=2.531$ previously seen in the QSO spectrum. If the redshift and the kinematics of both components of the double system are spectroscopically confirmed in the future, it would be an important high-z evidence of a merging system as the Lyman Limit absorber.

After carefully subtracting the PSF using two different PSF reconstruction approaches, we are able to see marginally the extended emission from the quasar host galaxy.  Since we simply and conservatively scaled the PSF peak flux to match the QSO central peak intensity and aligned them in both two approaches, we would obtain an oversubtracted extended emission from the quasar host, and can only place a low limit for the host flux.

Although the PSF subtracted residual images are mostly similar in the morphological structure for the two adopted PSF construction approaches , as well as among three consecutive observing nights, we noticed that different PSF subtraction algorithms would have large systematic errors on the photometry of the extended host emission. The systematic errors on the photometry of the extended host galaxy among three consecutive observing nights from Sept. 17-19, using the first approach of section 2.3, reach about one magnide, where PSF was constructed by averagely coadded interleaving exposures of PSF star. On the other hand, more reliable PSF reconstruction algorithm, such as PCA (the second approach of section 2.3), presented relatively consistent photometric measurement for the host galaxy among three consecutive observing nights, and possibly place a low limit of $m_{k}=23.3$ for the host galaxy (see detailed discussion in section 3.2). Further analyses and simulations on how to estimate properly the host properties are developing and will be presented by next work (He et al. in preparation).

\begin{acknowledgements}
This project/publication was made possible through the support of a grant from the John Templeton Foundation and National Astronomical Observatories of Chinese Academy of Sciences. The opinions expressed in this publication are those of the author(s) do not necessarily reflect the views of the John Templeton Foundation or National Astronomical Observatories of Chinese Academy of Sciences. The funds from John Templeton Foundation were awarded in a grant to The University of Chicago which also managed the program in conjunction with National Astronomical Observatories, Chinese Academy of Sciences. YPW would thank Dr. Chien Peng for the support on the GALFIT fitting process. YPW acknowledges the Subaru team and Dr. Yosuke Minowa for the hospitality and National Scientific Fundation of China(NSFC 10173025, 10673013 and 10778709) and the Chinese 973 project(TG 2000077602). We would like to thank our referee Dr. Michael Strauss for the helpful comments.
\end{acknowledgements}

\label{lastpage}

\end{document}